\documentclass[]{article}
\usepackage{lmodern}
\usepackage{amssymb,amsmath}

\IfFileExists{upquote.sty}{\usepackage{upquote}}{}
\IfFileExists{microtype.sty}{%
\usepackage{microtype}
\UseMicrotypeSet[protrusion]{basicmath} 
}{}
\PassOptionsToPackage{hyphens}{url}\usepackage{hyperref}
\hypersetup{unicode=true,
            pdftitle={Neighborhood Change, One Pint at a Time},
            pdfauthor={Jesus M. Barajas, Geoff Boeing, and Julie Wartell},
            pdfborder={0 0 0},
            breaklinks=true}
\usepackage{lscape}
\usepackage{longtable,booktabs}
\usepackage{graphicx,grffile}
\makeatletter
\def\maxwidth{\ifdim\Gin@nat@width>\linewidth\linewidth\else\Gin@nat@width\fi}
\def\maxheight{\ifdim\Gin@nat@height>\textheight\textheight\else\Gin@nat@height\fi}
\makeatother
\setkeys{Gin}{width=\maxwidth,height=\maxheight,keepaspectratio}
\IfFileExists{parskip.sty}{%
\usepackage{parskip}
}{
\setlength{\parindent}{0pt}
\setlength{\parskip}{6pt plus 2pt minus 1pt}
}
\ifx\paragraph\undefined\else
\let\oldparagraph\paragraph
\renewcommand{\paragraph}[1]{\oldparagraph{#1}\mbox{}}
\fi
\ifx\subparagraph\undefined\else
\let\oldsubparagraph\subparagraph
\renewcommand{\subparagraph}[1]{\oldsubparagraph{#1}\mbox{}}
\fi

\title{Neighborhood Change, One Pint at a Time: The Impact of Local Characteristics on Craft Breweries\footnote{Citation information: Barajas, J. M., Boeing, G., and Wartell, J. 2017. ``Neighborhood Change, One Pint at a Time: The Impact of Local Characteristics on Craft Breweries.'' In: \emph{Untapped: Exploring the Cultural Dimensions of Craft Beer}, edited by N. G. Chapman, J. S. Lellock, and C. D. Lippard. Morgantown, WV: West Virginia University Press. ISBN 978-1-943665-68-6.}}
\author{Jesus M. Barajas, Geoff Boeing, and Julie Wartell}
\date{}

\begin{document}
\maketitle

\begin{abstract}
Cities have recognized the local impact of small craft breweries, in many ways altering municipal codes to make it easier to establish breweries and making them the anchor points of economic development and revitalization. Nevertheless, we do not know the extent to which these strategies impacted changes at the neighborhood level across the nation. In this chapter, we examine the relationship between growth and locations of craft breweries and the incidence of neighborhood change across the United States. In the first part of the chapter, we rely on a unique dataset of geocoded brewery locations that tracks openings and closings from 2004 to the present. Using measures of neighborhood change often found in literature on gentrification-related topics, we develop statistical models relying on census tract demographic and employment data to determine the extent to which brewery locations are associated with social and demographic shifts since 2000. The strongest predictor of whether a craft brewery opened in 2013 or later in a neighborhood was the presence of a prior brewery. We do not find evidence entirely consistent with the common narrative of a link between gentrification and craft brewing, but we see a link between an influx of lower-to-middle income urban creatives and the introduction of a craft breweries. We advocate for urban planners to recognize the importance of craft breweries in neighborhood revitalization while also protecting residents from potential displacement.
\end{abstract}

\section{Introduction}\label{introduction}

Brooklyn Brewery occupies half a block along North 11th Street in the
heart of Williamsburg, one of New York City's most rapidly changing
neighborhoods. On most weekends, the tasting room is packed full of
enthusiastic craft beer drinkers. Former industrial spaces nearby house
gourmet restaurants, trendy bars, boutique hotels, and renovated
residential lofts. Williamsburg was not always this way. Remnants of the
area's industrial past are visible everywhere---large brick factory
buildings fill entire city blocks and a still-active oil depot operates
along the river inlet two blocks northwest of the brewery. Many consider
Brooklyn Brewery to be the anchor institution of Williamsburg's
revitalization, a popular narrative that, as we describe in this
chapter, repeats itself with Wynkoop Brewing in Denver's LoDo
neighborhood, 21st Amendment Brewery in San Francisco's SoMa
neighborhood, and others across the country.

Shifting consumer preferences toward more flavor, more options, and more
local products have fueled the growth of these three breweries and of
craft beer in general. However, urban planning and policy have also
influenced the success of craft brewing. Some cities have modified their
zoning regulations and offered financial incentives that have allowed
intrepid entrepreneurs to become first-movers into economically
uncertain locations (Best 2015; Hopkins 2014). In turn, these anchor
establishments helped spawn new, smaller craft breweries as the demand
for high-quality local beer---and other niche products and
services---has increased. Future growth of the craft beer industry is
tied to the success of these new breweries. Upstart brewers tend to be
small---often borne of a home brewing hobby---with the capacity and
profit incentives to serve only a local market. Unlike the pioneering
microbrewers before them that serve regional and multi-state consumers,
these newer brewers---such as brewpubs that produce beer only for the
customers who patronize their restaurants---require smaller production
spaces and thus are not limited to locating in industrial neighborhoods.
Seeking to capitalize on a new market of place-based consumers, newer
and smaller brewers may not be \emph{catalysts} of urban revitalization
so much as \emph{respondents} to changing neighborhood demographics.

In this chapter, we explore the influence of neighborhood change over
the past decade on where craft breweries are located. This study is the
first to empirically examine the relationship between neighborhoods and
craft breweries across the United States. Using US Census data, we first
describe the neighborhood characteristics of where craft breweries
operate. We then look at neighborhoods to understand how changes in
residential composition suggest factors influential in craft brewery
location decisions. We also explore differences at regional and
sub-regional spatial scales We conclude with some suggestions for urban
planning and policy as other cities turn to craft brewing as an
opportunity for neighborhood revitalization, economic development, and
tourism.

\section{Craft Beer: People, place, and
planning}\label{craft-beer-people-place-and-planning}

\subsection{People and place}\label{people-and-place}

Although the demand for craft beer has increased rapidly over the past
three decades, it has not grown uniformly across all demographic groups.
Craft beer drinkers tend to have higher incomes than other beer
drinkers, because on average craft beer commands a higher price than
other domestic or import beers (Tremblay and Tremblay 2011).
Furthermore, in the recent past, craft beer drinkers have tended to be
white, male generation X-ers (Tremblay and Tremblay 2005), but current
trends indicate an increasingly racially and ethnically diverse, female,
and millennial demographic profile (Watson 2014).

Several researchers have explored the link between place, demographics,
and the location of craft breweries. Three main threads link these
studies. First, the spatial geography of craft beer production in the
United States is uneven at multiple scales. At a regional level, for
example, the Pacific Coast states have seen major increases in
production volume and brewing facilities over the past three decades,
while there has been very little growth in the number of facilities in
the southeastern US over the same time period (McLaughlin, Reid, and
Moore 2014). To some extent, this is a result of California's vanguard
position in the rebirth of local brewing---many consider Anchor Steam in
San Francisco, New Albion in Sonoma, and Sierra Nevada in Chico to be
the modern founders of today's craft brewing industry (Acitelli 2013).
At the state level, research has found a significant association between
demographics and craft beer. Higher population predicts more craft beer
production, but controlling for population, traditional brewing culture
is a stronger predictor (McLaughlin, Reid, and Moore
201\protect\hypertarget{move422762242}{}{}4). Higher educational
attainment and greater levels of happiness and well-being may also be
associated with the amount of craft brewing at the state level (Florida
2012).

Cultural attitudes and affinities associated with place may impact craft
beer production and consumption in a metropolitan area. For example, the
values of residents who have helped to ``keep Portland weird'' may have
also contributed to the explosion of the craft beer industry in that
city (Cortright 2002). On the other hand, religious convictions and
corporate influence are significant predictors of the low number of
craft breweries in the southern United States (Gohmann 2015).
Metropolitan-level influences on craft beer also vary by region. Factors
such as the cost of living and the level of tolerance the population has
for activities outside of cultural norms are significant predictors of
craft breweries' presence in the South. Education levels and the amount
of arts and culture in the metropolitan area are not significant
predictors in the South, even though they are in other regions (Baginski
and Bell 2011).

A second thread of research on craft beer and place focuses on the idea
of local production. Cultural geographers have used the term
\emph{neolocalism} to describe the present-day phenomenon of the desire
for the local: preferring the mom-and-pop shop on Main Street to the
anonymity and sameness of the ``big box'' store (Flack 1997). Much as
wine connoisseurs travel to wineries to experience the \emph{terroir} of
a vintner's product, or foodies look to experience local flavors in new
restaurants or farmers markets, craft beer drinkers seek out the local
connection between their favorite beverage and the place where it was
brewed.

Many craft breweries tie into local landmarks and lore through their
beer names and labels. This can help newcomers share in the cultural
history of a place through consumption of a distinctively local product
(Schnell and Reese 2003), creating a common narrative of a certain
neighborhood history as new residents move in. For example, the Great
Lakes Brewing Company, based in Cleveland, Ohio, brews Burning River
Pale Ale, whose label pays tribute to the infamous 1969 Cuyahoga River
fire as a symbol of the city's industrial past and modern rebirth
(Stradling and Stradling 2008). Oakland's Linden Street Brewery ties
into the local ethos by delivering its flagship product in kegs solely
by bicycle to restaurants and bars in the city. The cargo bike that sits
in front of its brewing facility serves as a visible symbol of local
production and consumption, as well as its membership in the city's
bicycle culture.

A third rationale for the connection between craft beer and place can be
seen through the literature on gentrification. The term
\emph{gentrification} does not have a unique definition, but generally
refers to the process of middle-class professionals moving to
disinvested central city neighborhoods, upgrading housing, and
attracting new businesses that cater to the new neighborhood clientele.
Often, this process coincides with the displacement of current residents
and businesses, who tend to be poorer and from racial and ethnic
minority groups. Some have argued that as a result of America's
post-industrial economy, the newly-enlarged occupational class of
managers and technical professionals, usually considered the
gentrifiers, has had a substantial impact on consumer tastes and housing
preferences as they seek the culture and compactness of the central city
(e.g. Lees, Slater, and Wyly 2008).

Some scholars have argued that in certain locations, the development of
craft breweries can accelerate gentrification by playing on the
industrial heritage of the past (such as old manufacturing sites),
appealing to the ``discerning'' consumer class attracted to such
amenities, and in turn anchoring subsequent development (Mathews and
Picton 2014). \protect\hypertarget{move4227634296}{}{}In some respects,
then, craft beer is entangled with the process of neighborhood change,
and may be either a leading indicator (as a pioneer of reinvestment) or
a lagging indicator (as a response to changing tastes and local culture)
(e.g. Cortright 2002).

\subsection{Urban planning and policy}\label{urban-planning-and-policy}

Craft beer has become intertwined with city planning over the past
decade for two related reasons. First, it is increasingly seen as an
engine of local economic development and neighborhood vitality. Second,
for reasons articulated in the previous section, craft beer is readily
identified with its place of origin and attracts well-educated, affluent
consumers. As a result, civic leaders and city planners increasingly
look to the craft beer industry to play a role in neighborhood
revitalization (cf. Hackworth and Smith 2001). Efforts to revitalize
once-declining inner cities have emphasized the importance of the
``creative class'' and their demands in shaping reinvestment. Some have
argued that because creative professionals drive the new economy, cities
that wish to improve their economic performance should invest in the
amenities that attract this class of people (Florida 2002)---museums,
cultural activities, and perhaps craft beer.

Craft breweries are common first-movers into economically-depressed
neighborhoods often out of necessity. Larger breweries require expensive
equipment and ample space. Inexpensive rent is essential to keep
overhead costs low for this type of entrepreneurial light industry, much
as Jane Jacobs ({[}1961{]} 1992) noted in her praise of aged buildings.
Breweries may also produce unpleasant noise, odors, and considerable
wastewater pollution, making them less likely to obtain permits in
bedroom suburbs or upscale shopping centers. Nevertheless, with its
large equipment and high fixed costs, a new brewery signals that someone
is starting to invest long-term in a place, more so than does a bar or
restaurant. Smaller breweries may then follow. In turn, services move
in, young families begin to settle, a community grows, and craft
breweries become the canary in the coal mine for neighborhood change.

Significant anecdotal evidence suggests that this pattern is common. The
21st Amendment brewery in San Francisco is sometimes considered the
``granddaddy'' of the South of Market neighborhood (Associated Press
2013b), which is rapidly changing as a result of the region's technology
sector. The city of Oakland, California's senior economic development
specialist has argued that breweries revitalize struggling neighborhoods
by serving as a magnet for new businesses while creating foot traffic
and social activity (Somerville 2013).

Accordingly, city planners have recently begun to play an active role in
fostering brewery openings and the inchoate neighborhood revitalization
that trails them. Great Lakes Brewing opened in Cleveland's
economically-depressed Ohio City neighborhood in 1988. When it started
attracting customers and other shops began to open nearby, the city
repaved surrounding streets with cobblestones and invested millions of
dollars in the redevelopment of a neighboring abandoned historic market
hall (Associated Press 2013b). On the West Coast, Portland, Oregon's
development commission assists craft breweries with building renovations
(Best 2015). Since 2000, Portland has spent \$96 million on
revitalization efforts in its Lents neighborhood, recently dedicating
\$1 million to building improvements and loans for a new brewpub (Boddie
2014).

City planners also play a role through permitting and land use
regulation. Craft breweries' amalgam of industrial and retail uses often
necessitates special zoning, infrastructure, and government assistance
(Perritt 2013). Some cities, including San Diego, Long Beach, Dallas,
Charlotte, and Cincinnati, have introduced specific microbrewery land
use or more mixed-use designations to simplify development (City of San
Diego 2015; Appleton 2012; Peters and Szczepaniak 2013; May and Monk
2015). The San Francisco Brewers' Guild recently stated that the biggest
challenge facing their brewers was the long delay in acquiring permits
from the Department of Building Inspection, due to a permitting logjam
from the city's construction boom (Crowell 2013).

Today, many craft brewers explicitly view themselves as agents of
neighborhood revitalization and change (e.g., Bartlett et al. 2013;
Flynn et al. 2014). In 1989, Boulevard Brewing opened in central Kansas
City, Missouri. Rather than locating on inexpensive land at the urban
periphery, the brewers wanted to contribute to the central city's urban
vitality, referring to themselves as ``committed urbanists'' (Associated
Press 2013a). However, urban breweries like Boulevard may sometimes
become victims of their own success; the ensuing desire to be in the
neighborhood can increase local rents and demand for space. Marquee
breweries such as 21st Amendment and Brooklyn Brewery have begun to
expand and relocate to less-expensive neighborhoods in their
metropolitan areas (Associated Press 2013b; Li 2014), perhaps restarting
the cycle of neighborhood change elsewhere, anew.

The news media, case studies, and development reports provide many of
the details behind the effects of craft breweries and the economic
transitions of neighborhoods in which they are located. In summary,
craft brewing is good for cities by investing in struggling
neighborhoods and adding an amenity to changing neighborhoods, and
planners are willing to accommodate these new investments. Much of the
information on a sub-metropolitan scale relies on anecdotal evidence or
single case studies. There is little empirical evidence from these
studies or others that assess the relationships between residential
characteristics of neighborhoods and craft breweries.

\section{Methods and Data}\label{methods-and-data}

Craft brewing is related to neighborhood change and urban planning, but
the nature of this relationship remains vague in the research
literature. To address this, we explored the extent to which the
changing residential characteristics of neighborhoods influence the
location of new breweries. In other words, does urban revitalization
predict the locations of these desirable assets? Given the
cross-sectional nature of our dataset, it is not possible to assign a
direction of causality to the relationship between craft brewery
locations and neighborhood change---and, in fact, there may be a
reciprocal relationship. Nevertheless, understanding associations
between craft brewing and neighborhood change has policy implications,
such as whether cities should create incentives for breweries to locate
in disinvested neighborhoods if demographic changes encourage them to
locate in revitalized neighborhoods otherwise.

\subsection{Craft brewery locations}\label{craft-brewery-locations}

To understand the relationship between craft brewing and location, we
obtained a unique dataset of craft breweries in the United States from
PubQuest (2015), a company that maps craft breweries.\footnote{The third
  author of this paper is the co-founder of PubQuest.} PubQuest compiles
the brewery data from a variety of public and private sources and
includes each US craft brewery location open to the public, the address,
and the type of brewery. Brewery types include brewery with tasting
room, brewpub with on-site brewing and food service, and brew houses
that are owned by craft breweries without on-site brewing. The dataset
includes all craft breweries that were in operation at some time between
2006 and 2015. The dataset does not include production volume, though
all breweries listed meet the Brewers Association definition of a craft
brewer.

We aggregated the PubQuest data to the census tract level to harmonize
with US Census socioeconomic variables. Census tracts are an imperfect
spatial unit and may miss more localized relationships between
neighborhood change and craft brewery locations. Nevertheless, they
provide a consistent level of geography across the United States and are
relatively stable over time (though see discussion below). Socioeconomic
data are more reliable at the census tract level than at smaller spatial
units, such as block groups. Thus, we defined neighborhoods using the
census tract as the spatial unit.

\subsection{Identifying neighborhood
change}\label{identifying-neighborhood-change}

We measured change between the 2000 decennial Census and the 2009--2013
five-year American Community Survey (ACS) estimates. The ACS aggregates
survey responses from each year of the five-year period into one
dataset, so it allows for a rough comparison of 2000 and 2011
socioeconomic data. Literature on gentrification and displacement
provided a starting point for variables appropriate to measure when
trying to understand neighborhood change (e.g. Freeman 2005; Newman and
Wyly 2006). We selected variables on race and ethnicity, age, family
structure, educational attainment, income, employment, housing age,
median home value, and population density as independent variables. All
dollar amounts are inflation-adjusted 2013 dollars.

Although we are primarily interested in how change in these variables is
associated with craft brewery locations, we also included year 2000
values for each variable for which we examined change. In this way, we
controlled for locations that may have experienced little change but had
high or low values of each variable to begin with. We standardized
census tract definitions to the 2010 boundaries, using the Brown
University Longitudinal Tract Database files (Logan, Xu, and Stults
2014). We only included variables in our models that are strictly
comparable between the two datasets.

We estimated a series of logistic regression models, in which the
dependent variable is whether a census tract has a new craft brewery. We
estimated both standard and robust versions of the models. We defined
``new'' to mean whether a brewery opened in 2013 or later. In each
model, we controlled for whether a craft brewery existed prior to 2013.
Particularly in the early years of the craft brewing renaissance,
brewers relied on existing knowledge of those who came before them in
the industry (Ogle 2006; Acitelli 2013), so we expected to see a
positive relationship between the presence of an older brewery and a new
one. We also controlled for the census division in which the tract is
located, as defined by the US Census Bureau (2000). As we described
earlier, existing research has shown an uneven geographic distribution
of craft beer, with far greater prevalence in the western US and far
less prevalence in the southeastern portion of the country (Baginski and
Bell 2011; McLaughlin, Reid, and Moore 2014; Grohmann 2015). Independent
variables in the models are cross-sectional and paired together: one
member of the pair measures change over time while the other measures
the value in the base year.

\section{Craft Beer and Neighborhood
Change}\label{craft-beer-and-neighborhood-change}

\subsection{Descriptive statistics}\label{descriptive-statistics}

\begin{figure}
	\centering
		\includegraphics[width=0.95\linewidth]{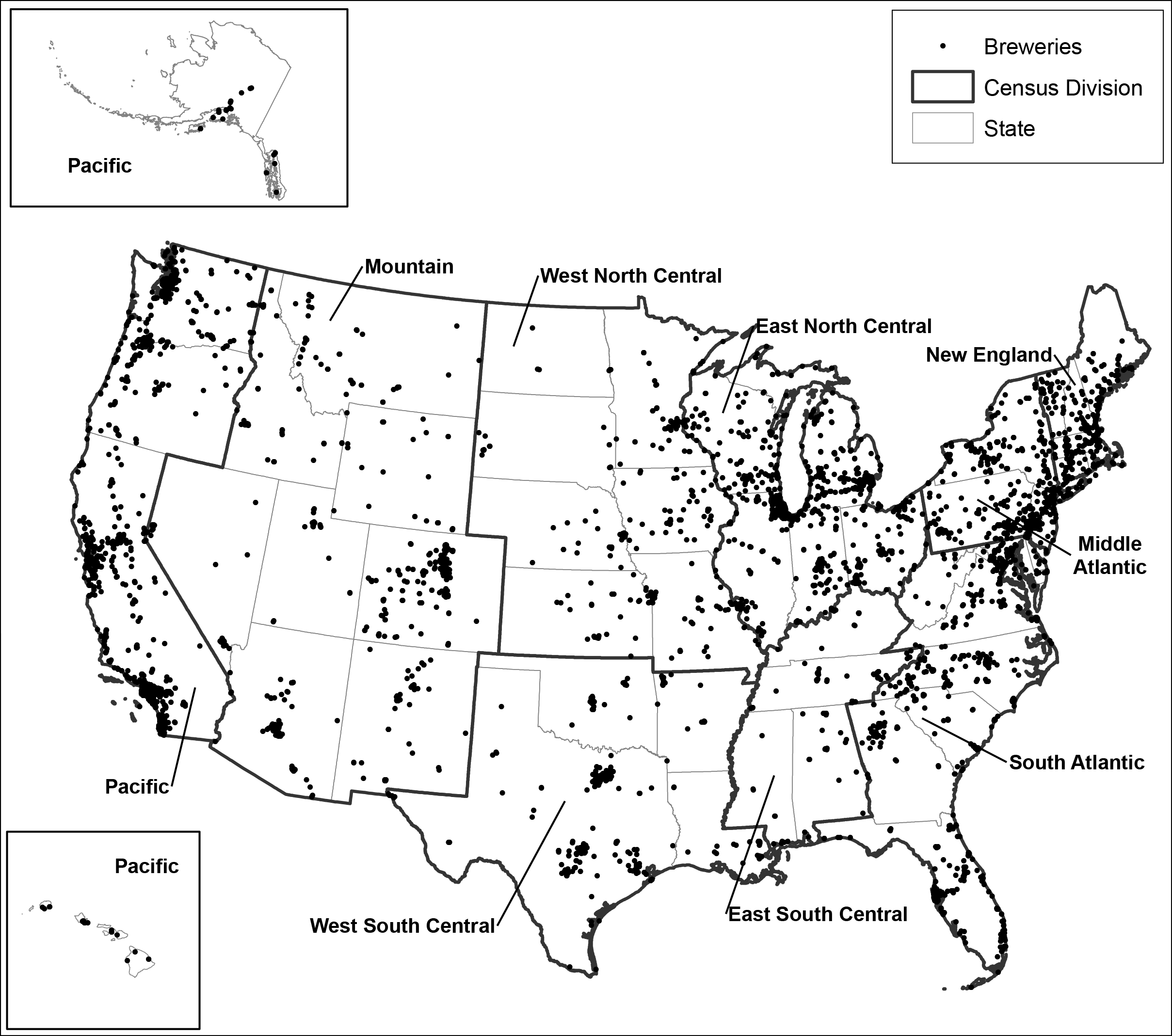}
	\caption{Craft breweries are predominately located west of the Rocky
Mountains. There are a significant number of craft breweries in
metropolitan areas of the Midwest and Northeast. Far fewer are located
in the southern United States. Labels indicate census divisions. Data
sources: PubQuest (2015), US Census Bureau (2000).}
	\label{fig:usa}
\end{figure}

\begin{figure}
	\centering
		\includegraphics[width=0.95\linewidth]{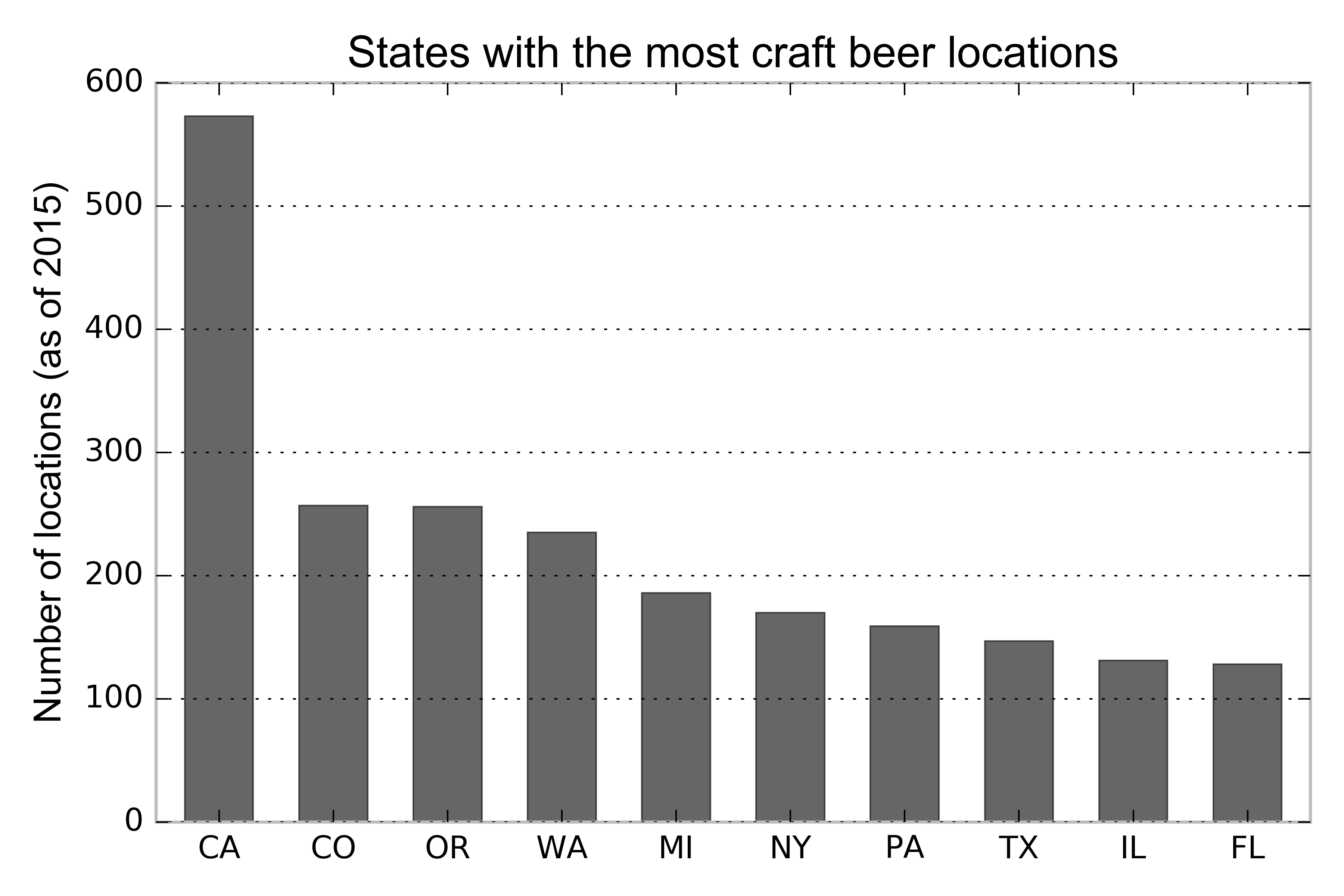}
	\caption{Craft brewery locations by state. In March 2015, California had 573
operating craft breweries, more than double that of the next state,
Colorado. Even though the western US dominates craft brewing, six of the
top ten states with the most craft breweries are in the eastern and
southern US. Data source: PubQuest (2015).}
	\label{fig:breweries}
\end{figure}

As of March 2015, 4,044 craft brewery locations as we defined them were
in operation in the United States, approximately half of which (2,036)
had opened in 2013 or later. The Pacific census division has the most
craft breweries with 884, while the East South Central division, which
includes Alabama, Kentucky, Mississippi, and Tennessee, has the fewest
with 73 (Figures \ref{fig:usa} and \ref{fig:breweries}). Most of the Pacific's breweries are in
California, which has more than twice the number of craft breweries
compared to the next largest concentration in Colorado. Several states
outside the West, including Michigan, New York, and Pennsylvania, also
have a significant number of breweries. Breweries are not clustered at
the neighborhood level across the US---fewer than 4\% of census tracts
had one or more breweries, and 88\% of those had only one. However, over
90\% of census tracts with craft breweries are in urbanized areas or
urban clusters as defined by the Census Bureau, which means that almost
all breweries are near concentrations of at least 2,500 people.

Mean values for several demographic, housing, and employment variables
using 2009--2013 ACS estimates are shown in Table 1, categorized
according to whether the census tract had at least one craft brewery
location in 2015. Neighborhoods with craft breweries are about
two-thirds as dense as those without (equivalent to densities of many
inner-ring suburbs). Those neighborhoods tend to have more white
residents, a higher proportion of people in the 25--34 year age range,
fewer households with children, more education, a slightly lower median
income, a higher proportion of people in professional or technical
occupations, more new housing, and higher housing prices.

\begin{longtable}[c]{@{}lrr@{}}
\caption{Mean values per census tract}\\
\toprule
& With craft brewery & Without craft brewery\tabularnewline
\midrule
\endhead
Population density (sq. mi.) & 3,604 & 5,377\tabularnewline
Non-Hispanic White & 71\% & 63\%\tabularnewline
Black/African American & 8\% & 14\%\tabularnewline
Hispanic/Latino & 13\% & 16\%\tabularnewline
Age 24 or younger & 32\% & 33\%\tabularnewline
Age 25-34 & 16\% & 13\%\tabularnewline
\emph{Age 35-44} & \emph{13\%} & \emph{13\%}\tabularnewline
Age 45-54 & 14\% & 14\%\tabularnewline
\emph{Age 55-64} & \emph{12\%} & \emph{12\%}\tabularnewline
Age 65 or older & 14\% & 14\%\tabularnewline
Households with children & 27\% & 33\%\tabularnewline
Earned bachelor's degree & 34\% & 27\%\tabularnewline
Median income (2013\$) & \$54,585 & \$56,834\tabularnewline
Employed & 90\% & 90\%\tabularnewline
Professional occupations & 62\% & 59\%\tabularnewline
Housing built since 2000 & 15\% & 13\%\tabularnewline
Median housing age (yrs) & 43 & 42\tabularnewline
Median house price (2013\$) & \$250,046 & \$217,988\tabularnewline
\bottomrule
\multicolumn{3}{l}{Note: All differences significant at 95\% confidence interval except} \\
\multicolumn{3}{l}{those in italics. Data source: 2009-2013 5-year ACS estimates.}
\end{longtable}

\subsection{The effects of neighborhood change on new
breweries}\label{the-effects-of-neighborhood-change-on-new-breweries}

Analytical results are shown in Table 2. Coefficients and significance
levels in both the standard and robust versions of the models were
similar and so their interpretations did not change substantially. Thus,
we show only the standard model estimates. Values in the table represent
odds ratios; that is, the odds that there is a new brewery in the tract
compared to the odds that there is not, given the presence of the
variable. Each of the first three models tests the influence of a
different set of indicators on the location of new craft breweries.

The first model tests the influence of racial and ethnic categories.
Breweries were more likely to locate in census tracts that lost racial
and ethnic diversity over the decade we examined. Taken together, higher
proportions of each of the racial and ethnic groups in the year 2000
were statistically significantly associated with a brewery opening.
However, declines in the black and Latino populations between 2000 and
the 2009--2013 ACS estimate predicts greater odds of a new craft brewing
location. For each percentage point decline in the black and Latino
populations, breweries were about three percent more likely to open.

Model 2 tests the relationship between other socioeconomic indicators
and craft breweries. The results give some credence to the idea that
breweries are locating in places where they can cater to a younger urban
professional crowd. An increase in the 25-to-34 year old population and
the proportion of older residents is positively associated with craft
brewery locations. Fewer households without children in 2000 is also
statistically significant. Both high levels of college-educated
residents and increases in the proportion of people with college degrees
predict the opening of a brewery, while the relationship is the inverse
for income level. We hypothesized that craft brewing would appeal to a
professionally-employed population. However, change in this occupational
classification is statistically insignificant. Lower proportions of
residents who were professionally employed in 2000 are associated with
presence of neighborhood breweries.

The third model tests the influence of housing factors on craft brewery
locations, primarily as a proxy for urban growth and change. We find
that areas with a higher proportion of new housing stock in 2000, an
increasing proportion of newly-built housing, and a decline in the
median housing age predict craft brewing. At the same time, each year
older the median housing age is in the census tract, the odds of a new
craft brewery location increase by about 3.6\%. The model indicates that
census tracts with higher home values are likely to have breweries,
although the change in home values is insignificant. We suspect this
could mean that breweries are locating in census tracts with more infill
development rather than those with expansive growth, indicating brewers'
preferences for urban or central locations rather than outlying
locations.

Model 4 tests the simultaneous influence of all three categories on
craft brewery openings. To reduce issues of multicollinearity, we
removed variables with variance inflation factors greater than 7. Most
of the variables from the first three models remain significant, with a
few notable differences. The black population in a census tract is no
longer a significant predictor of breweries, though an increase in the
white population becomes significant. Lower median incomes in a
neighborhood have stronger effects on a brewery opening compared to the
models that do not control for race and housing variables.

In all model specifications, we controlled for whether the census tract
had a craft brewery prior to 2013, the population density of the census
tract, and the geographic region of the country. The effect of a
previous brewery is strong: the odds of a new craft brewery opening in a
census tract are 2.8 times greater if there had been one prior, when
controlling for all other characteristics. The result suggests older
breweries act as catalysts for new breweries to co-locate. Population
density has a small but negative relationship with brewery locations:
adding one hundred additional people per square mile reduces the odds of
a new brewery by about one percent. Consistent with other research
(Baginski and Bell 2011), we found that the odds are significantly
higher for breweries to open in the western US and significantly lower
in the southern US.

\begin{landscape}
\begin{longtable}[c]{@{}lccccc@{}}
  \caption{Logistic regression models of neighborhood change impacts on breweries. } \\
\toprule
 & (1) & (2) & (3) & (4) & (5)\\ 
 & Race/ethnicity & Other SES & Housing & Full model & West Coast \\
\midrule
\endfirsthead

\toprule
 & (1) & (2) & (3) & (4) & (5)\\ 
 & Race/ethnicity & Other SES & Housing & Full model & West Coast \\
\midrule
\endhead

 Non-Hispanic White (\%) & 1.017$^{***}$ &  &  & 1.010$^{***}$ & 1.012$^{**}$ \\ 
  & (1.010, 1.025) &  &  & (1.006, 1.014) & (1.001, 1.023) \\ 
  & & & & & \\ 
 Change in non-Hispanic White (pp) & 1.001 &  &  & 1.013$^{***}$ & 1.011 \\ 
  & (0.990, 1.012) &  &  & (1.005, 1.021) & (0.991, 1.032) \\ 
  & & & & & \\ 
 Black/African American (\%) & 1.012$^{***}$ &  &  &  &  \\ 
  & (1.005, 1.020) &  &  &  &  \\ 
  & & & & & \\ 
 Change in Black/African American (pp) & 0.970$^{***}$ &  &  &  &  \\ 
  & (0.957, 0.983) &  &  &  &  \\ 
  & & & & & \\ 
 Hispanic or Latino (\%) & 1.013$^{***}$ &  &  & 1.013$^{***}$ & 1.012$^{*}$ \\ 
  & (1.006, 1.021) &  &  & (1.008, 1.018) & (1.000, 1.026) \\ 
  & & & & & \\ 
 Change in Hispanic or Latino (pp) & 0.977$^{***}$ &  &  & 1.010$^{*}$ & 1.005 \\ 
  & (0.965, 0.989) &  &  & (0.999, 1.020) & (0.983, 1.028) \\ 
  & & & & & \\ 
 Age 25 or younger (\%) &  & 1.000 &  &  &  \\ 
  &  & (0.989, 1.012) &  &  &  \\ 
  & & & & & \\ 
 Change in age 25 or younger (pp) &  & 0.997 &  &  &  \\ 
  &  & (0.981, 1.013) &  &  &  \\ 
  & & & & & \\ 
 Age 25-34 (\%) &  & 0.998 &  & 0.995 & 1.037$^{**}$ \\ 
  &  & (0.982, 1.014) &  & (0.980, 1.010) & (1.001, 1.075) \\ 
  & & & & & \\ 
 Change in age 25-34 (pp) &  & 1.019$^{**}$ &  & 1.031$^{***}$ & 1.006 \\ 
  &  & (1.000, 1.038) &  & (1.013, 1.050) & (0.965, 1.050) \\ 
  & & & & & \\ 
 Age 35-44 (\%) &  & 1.088$^{***}$ &  & 1.107$^{***}$ & 1.071$^{**}$ \\ 
  &  & (1.057, 1.121) &  & (1.081, 1.134) & (1.016, 1.129) \\ 
  & & & & & \\ 
 Change in age 35-44 (pp) &  & 0.925$^{***}$ &  & 0.914$^{***}$ & 0.951$^{*}$ \\ 
  &  & (0.898, 0.952) &  & (0.894, 0.936) & (0.903, 1.002) \\ 
  & & & & & \\ 
 Age 45-54 (\%) &  & 1.036$^{**}$ &  &  &  \\ 
  &  & (1.003, 1.070) &  &  &  \\ 
  & & & & & \\ 
 Change in age 45-54 (pp) &  & 0.963$^{**}$ &  &  &  \\ 
  &  & (0.932, 0.995) &  &  &  \\ 
  & & & & & \\ 
 Age 55-64 (\%) &  & 0.873$^{***}$ &  & 0.888$^{***}$ & 0.921$^{**}$ \\ 
  &  & (0.836, 0.912) &  & (0.864, 0.914) & (0.860, 0.987) \\ 
  & & & & & \\ 
 Change in age 55-64 (pp) &  & 1.136$^{***}$ &  & 1.124$^{***}$ & 1.103$^{***}$ \\ 
  &  & (1.093, 1.180) &  & (1.090, 1.159) & (1.024, 1.189) \\ 
  & & & & & \\ 
 Households with children (\%) &  & 0.953$^{***}$ &  & 0.956$^{***}$ & 0.955$^{***}$ \\ 
  &  & (0.945, 0.961) &  & (0.948, 0.964) & (0.935, 0.974) \\ 
  & & & & & \\ 
 Change in household with children (pp) &  & 0.994 &  & 0.994 & 1.004 \\ 
  &  & (0.985, 1.003) &  & (0.985, 1.003) & (0.984, 1.025) \\ 
  & & & & & \\ 
 Earned bachelors (\%) &  & 1.026$^{***}$ &  & 1.022$^{***}$ & 1.014 \\ 
  &  & (1.019, 1.033) &  & (1.015, 1.029) & (0.996, 1.032) \\ 
  & & & & & \\ 
 Change in earned bachelors (pp) &  & 1.030$^{***}$ &  & 1.026$^{***}$ & 1.021$^{**}$ \\ 
  &  & (1.022, 1.039) &  & (1.017, 1.035) & (1.000, 1.043) \\ 
  & & & & & \\ 
 Log of median income (2013\$) &  & 0.426$^{***}$ &  & 0.293$^{***}$ & 0.404$^{**}$ \\ 
  &  & (0.332, 0.546) &  & (0.218, 0.394) & (0.196, 0.828) \\ 
  & & & & & \\ 
 Change in median income (\%) &  & 0.994$^{***}$ &  & 0.992$^{***}$ & 0.988$^{***}$ \\ 
  &  & (0.991, 0.996) &  & (0.990, 0.995) & (0.981, 0.995) \\ 
  & & & & & \\ 
 Employed (\%) &  & 1.011 &  & 0.999 & 1.060$^{***}$ \\ 
  &  & (0.998, 1.024) &  & (0.984, 1.015) & (1.021, 1.102) \\ 
  & & & & & \\ 
 Change in employed (pp) &  & 1.013$^{**}$ &  & 1.004 & 1.030$^{**}$ \\ 
  &  & (1.002, 1.024) &  & (0.993, 1.016) & (1.003, 1.058) \\ 
  & & & & & \\ 
 Professional occupations (\%) &  & 0.974$^{***}$ &  & 0.979$^{***}$ & 0.960$^{***}$ \\ 
  &  & (0.966, 0.983) &  & (0.971, 0.988) & (0.939, 0.981) \\ 
  & & & & & \\ 
 Change in professional occupations (pp) &  & 1.002 &  & 1.000 & 1.009 \\ 
  &  & (0.994, 1.009) &  & (0.992, 1.008) & (0.992, 1.027) \\ 
  & & & & & \\ 
 Housing built in last 10 years (\%) &  &  & 1.020$^{***}$ & 1.013$^{***}$ & 1.008 \\ 
  &  &  & (1.015, 1.025) & (1.007, 1.018) & (0.995, 1.021) \\ 
  & & & & & \\ 
 Change in housing built in last 10 years (pp) &  &  & 1.011$^{***}$ & 1.005$^{**}$ & 0.997 \\ 
  &  &  & (1.007, 1.016) & (1.000, 1.010) & (0.985, 1.008) \\ 
  & & & & & \\ 
 Median housing age (yrs) &  &  & 1.036$^{***}$ & 1.014$^{***}$ & 1.004 \\ 
  &  &  & (1.031, 1.042) & (1.008, 1.019) & (0.990, 1.017) \\ 
  & & & & & \\ 
 Change in median housing age (\%) &  &  & 0.999$^{*}$ & 1.000 & 0.999 \\ 
  &  &  & (0.998, 1.000) & (0.998, 1.001) & (0.995, 1.001) \\ 
  & & & & & \\ 
 Log of median home value (2013\$) &  &  & 1.152$^{***}$ & 1.570$^{***}$ & 1.251 \\ 
  &  &  & (1.052, 1.261) & (1.323, 1.865) & (0.883, 1.786) \\ 
  & & & & & \\ 
 Change in median home value (\%) &  &  & 1.003 & 1.006$^{*}$ & 1.008 \\ 
  &  &  & (0.991, 1.008) & (0.999, 1.012) & (0.993, 1.017) \\ 
  & & & & & \\ 
 Brewery before 2013 (1 = yes) & 5.754$^{***}$ & 3.020$^{***}$ & 5.265$^{***}$ & 2.790$^{***}$ & 3.317$^{***}$ \\ 
  & (4.977, 6.632) & (2.578, 3.527) & (4.540, 6.085) & (2.373, 3.268) & (2.490, 4.386) \\ 
  & & & & & \\ 
 Population density (sq. mi.) & 1.000$^{***}$ & 1.000$^{***}$ & 1.000$^{***}$ & 1.000$^{***}$ & 1.000$^{***}$ \\ 
  & (1.000, 1.000) & (1.000, 1.000) & (1.000, 1.000) & (1.000, 1.000) & (1.000, 1.000) \\ 
  & & & & & \\ 
 Mid Atlantic & 0.594$^{***}$ & 0.696$^{***}$ & 0.606$^{***}$ & 0.769$^{**}$ &  \\ 
  & (0.473, 0.747) & (0.553, 0.876) & (0.482, 0.762) & (0.609, 0.972) &  \\ 
  & & & & & \\ 
 South Atlantic & 0.574$^{***}$ & 0.502$^{***}$ & 0.752$^{**}$ & 0.646$^{***}$ &  \\ 
  & (0.464, 0.712) & (0.405, 0.624) & (0.604, 0.940) & (0.513, 0.816) &  \\ 
  & & & & & \\ 
 East North Central & 0.818$^{*}$ & 0.800$^{**}$ & 0.864 & 0.894 &  \\ 
  & (0.670, 1.003) & (0.652, 0.985) & (0.705, 1.064) & (0.726, 1.107) &  \\ 
  & & & & & \\ 
 East South Central & 0.307$^{***}$ & 0.249$^{***}$ & 0.410$^{***}$ & 0.321$^{***}$ &  \\ 
  & (0.219, 0.424) & (0.176, 0.347) & (0.289, 0.572) & (0.225, 0.452) &  \\ 
  & & & & & \\ 
 West North Central & 0.644$^{***}$ & 0.558$^{***}$ & 0.687$^{***}$ & 0.660$^{***}$ &  \\ 
  & (0.505, 0.821) & (0.435, 0.716) & (0.535, 0.883) & (0.510, 0.853) &  \\ 
  & & & & & \\ 
 West South Central & 0.415$^{***}$ & 0.361$^{***}$ & 0.492$^{***}$ & 0.460$^{***}$ &  \\ 
  & (0.319, 0.540) & (0.275, 0.471) & (0.372, 0.648) & (0.343, 0.614) &  \\ 
  & & & & & \\ 
 Mountain & 1.255$^{**}$ & 1.201 & 1.624$^{***}$ & 1.206 &  \\ 
  & (1.009, 1.566) & (0.962, 1.504) & (1.294, 2.044) & (0.952, 1.533) &  \\ 
  & & & & & \\ 
 Pacific & 1.291$^{**}$ & 1.349$^{***}$ & 1.396$^{***}$ & 1.350$^{***}$ &  \\ 
  & (1.047, 1.598) & (1.101, 1.661) & (1.141, 1.717) & (1.089, 1.680) &  \\ 
\midrule
Observations & 72,815 & 72,518 & 71,205 & 71,204 & 9,980 \\ 
Log Likelihood & $-$8,114.107 & $-$7,682.520 & $-$7,928.172 & $-$7,518.136 & $-$1,371.342 \\ 
Akaike Inf. Crit. & 16,262.210 & 15,427.040 & 15,890.340 & 15,110.270 & 2,800.683 \\ 
\bottomrule
\multicolumn{6}{l}{\textit{Note:} $^{*}$p$<$0.1; $^{**}$p$<$0.05; $^{***}$p$<$0.01} \\ 
\multicolumn{6}{l}{pp = percentage points. Values in parentheses are 95\% confidence intervals. Models 1-4 use nationwide data, } \\
\multicolumn{6}{l}{model 5 is the West Coast model.}
\end{longtable}
\end{landscape}

\subsection{A closer look: Brewing on the West
Coast}\label{a-closer-look-brewing-on-the-west-coast}

Because of the distinctive regional variation in craft brewing locations
and the role of pioneers in the western United States in establishing
the new craft brewing movement, we re-estimated our national model for
California, Oregon, and Washington. These three states contain about a
quarter of all craft breweries in the US. We expected that, in addition
to its unique history with respect to the craft brewing movement,
relatively higher living costs compared to the rest of the country might
change some of the relationships between neighborhoods and breweries.
Results are shown in Table 2 as model 5.

Fewer relationships are significant in the regional model compared to
the national model. An increase in the proportion of people aged 55--64
and people employed predict new breweries in a neighborhood. Curiously,
in both the West Coast model and the national model, an increase in the
number of bachelor's degrees and a decrease in median income are
associated with new brewery openings. We discuss this paradoxical
finding further in the discussion. Additionally, as in the national
model, the presence of an older brewery remains a significant predictor
of whether a new one opened in the previous two years. The results
suggest a smaller influence on neighborhood change factors in craft
brewing locations along the West Coast. It is possible that new craft
breweries are unable to open in significantly changing residential areas
because of other development pressures.

\section{Discussion and Conclusion}\label{discussion-and-conclusion}

Today's wave of urban revitalization efforts has been viewed by
supporters as a way to increase a city's wealth and economic
opportunities. Detractors, however, consider it gentrification with
better marketing. There has been considerable anecdotal evidence that
craft breweries are harbingers and even instigators of neighborhood
change. Many cities today pursue craft breweries as potential job
creators, catalysts for investment and development, and tourist
attractors. Craft beer consumption is associated with higher
socioeconomic indicators such as race, income, and education, (Tremblay
and Tremblay 2005; Tremblay and Tremblay 2011; Florida 2012) and, at
larger spatial scales, population growth and diversification (Schnell
and Reese 2014). This study is unique in that we looked at
neighborhood-level characteristics, which better reflect the character
of small craft brewers than regional- or state-level analyses can.

At a neighborhood level, we find a slightly different story than that
told by previous researchers. Our data do not allow us to conclude that
new craft breweries cause changes in neighborhood indicators, but we can
see how they follow change. Our results are not entirely consistent with
the story of gentrification in neighborhoods; changes in racial
composition do not seem to be a draw for new craft breweries, nor do
so-called ``creative class'' occupations. On the other hand, areas that
have a highly educated population, increasing education levels, lower
and declining income levels, and an older but developing housing stock
do appear to welcome craft brewing to the neighborhood. Perhaps we might
speak of a link between craft beer and what one writer has recently
called ``yuccies''---young, urban creatives, who may not fit in the
standard US Census employment categories (Infante 2015). These changes
also depend on the region, so influences in one may not necessarily be
significant in another. Our informal survey with 15 brewery owners
tentatively confirms these findings. They uniformly stated that
neighborhood character was very important or even the primary reason for
their location choice. Many referred to themselves explicitly as
pioneers and catalysts in neglected historic neighborhoods.

If craft breweries and brewpubs are paths to coveted neighborhood
revitalization, planners must keep several things in mind. First,
simplifying the permitting process and creating dedicated craft brewery
land use designations can reduce some of the bureaucratic obstacles to
the development of breweries. Such cities and neighborhoods thus become
more attractive. Second, coordinating revitalization efforts and
subsidies with potential breweries can create synergies in the
improvement of neighborhood infrastructure. Renovating buildings and
improving streets and sidewalks can maximize the effect of surrounding
economic development.

Future research may help strengthen some of these conclusions. For
example, we expect employment-side factors and planning regulations to
influence the locations of new craft breweries, there is no nationwide
dataset that would allow us to include those factors in our analysis at
the census tract level. For that reason, we only investigated the
influence of residential patterns on brewing locations and leave
investigations that include these factors at the subregional level for
future work. We also suspect that the effect of changes may be different
for different brewery types, but we leave this analysis for future work
as well. Finally, controlling for metropolitan differences in housing
costs might clarify the relationship between housing values and
investment and craft brewery locations.

City planners are also agents of neighborhood change and bear a
responsibility to current residents to represent their interests. Simply
allowing rents to rise as trendy businesses and affluent residents
arrive is not a good faith effort to represent the needs of longtime
residents. Rather, diverse and inclusive collaboration among all
impacted stakeholders is critical for equitable planning. How can local
culture and history be preserved while increasing economic opportunity
and amenities for all? Displacement of longtime residents is a key
challenge facing economic revitalization of disinvested neighborhoods.
If craft beer is a canary in the coal mine for neighborhood change,
perhaps it can also be a trigger for proactive planning interventions,
harnessing the image of the ``local'' to ensure people who made the
history in the images can remain in their place on the bar stool.

\clearpage
\newpage
\section{References}\label{references}

Acitelli, Tom. 2013. \emph{The Audacity of Hops: The History of
America's Craft Beer Revolution}. Chicago: Chicago Review Press.

Appleton, R. 2012. ``Council Relaxes Zoning Rules for Alcohol
Production.'' \emph{The Dallas Morning News}, June 27.
\url{http://cityhallblog.dallasnews.com/2012/06/council-relaxes-zoning-rules-for-alcohol-production-fourth-brewery-seeking-city-approval.html/}

Associated Press. 2013a. ``A Tale of 6 Cities Craft Brewers Helped
Transform.'' \emph{The Street}, July 4.
\url{http://www.thestreet.com/story/11969867/1/a-tale-of-6-cities-craft-brewers-helped-transform.html}

Associated Press. 2013b. ``How Craft Beer Is Reviving Urban
Neighborhoods.'' \emph{Business Insider}, July 4.
\url{http://www.businessinsider.com/craft-brews-create-urban-revival-2013-7}

Baginski, James, and Thomas L. Bell. 2011. ``Under-Tapped?: An Analysis
of Craft Brewing in the Southern United States.'' \emph{Southeastern
Geographer} 51 (1): 165--85. doi:10.1353/sgo.2011.0002.

Bartlett, Dan, Steve Allen, Jack Harris, and Larry Cary. 2013.
``Revitalization One Pint at a Time: How Breweries and Distilleries
Contribute to Main Street.'' Presented at the Oregon Main Street
Conference, Astoria, OR, October 2-4.
\url{https://www.oregon.gov/oprd/HCD/SHPO/docs/2013OMSConf/Revitalization\_By\_The\_Pint.pdf}

Best, Allen. 2015. ``Welcome to Beer Country.'' \emph{Planning},
February. \url{https://www.planning.org/planning/open/2015/welcometobeer.htm.}

Boddie, K. 2014. ``New Lents Brewers `Have a Track Record.''' \emph{KOIN
News}, March 18.
\url{http://koin.com/2014/03/18/new-lents-brewers-track-record/}

City of San Diego. 2015. \emph{Land Development Code}.
\url{http://www.sandiego.gov/development-services/industry/landdevcode/}

Cortright, Joseph. 2002. ``The Economic Importance of Being Different:
Regional Variations in Tastes, Increasing Returns, and the Dynamics of
Development.'' \emph{Economic Development Quarterly} 16 (1): 3--16.
doi:10.1177/0891242402016001001.

Crowell, C. 2013. ``Better Know a Craft Beer Guild: San Francisco Booms
amid Permit Delays.'' \emph{Craft Brewing Business}, August 22.
\url{http://www.craftbrewingbusiness.com/business-marketing/better-know-a-craft-beer-guild-san-francisco/}

Flack, Wes. 1997. ``American Microbreweries and Neolocalism: `Ale-Ing'
for a Sense of Place.'' \emph{Journal of Cultural Geography} 16 (2):
37--53. doi:10.1080/08873639709478336.

Florida, Richard. 2012. ``The Geography of Craft Beer.'' \emph{CityLab}.
\url{http://www.citylab.com/design/2012/08/geography-craft-beer/2931/.}

Florida, Richard L. 2002. \emph{The Rise of the Creative Class: And How
It's Transforming Work, Leisure, Community and Everyday Life.} New York:
Basic Books.

Flynn, Edward, Willard Brooks, Jeff Ware, Mark Woodcock, and Ted Hawley.
2014. ``Brewing Economic Development.'' Presented at the New York
Upstate Planning Association Conference, Rochester, NY, September 17-19.
\url{http://nyupstateplanning.org/wp-content/uploads/2014/10/6B-Brewing-Economic-Development.pdf}

Freeman, Lance. 2005. ``Displacement or Succession? Residential Mobility in Gentrifying Neighborhoods.'' \emph{Urban Affairs Review} 40 (4): 463--491. doi:10.1177/1078087404273341.

Gohmann, Stephan F. 2015. ``Why Are There so Few Breweries in the
South?'' \emph{Entrepreneurship Theory and Practice}.
doi:10.1111/etap.12162.

Hackworth, Jason, and Neil Smith. 2001. ``The Changing State of
Gentrification.'' \emph{Tijdschrift Voor Economische En Sociale
Geografie} 92 (4): 464--77. doi:10.1111/1467-9663.00172.

Hopkins, David. 2014. ``How Craft Beer (Finally) Came to Dallas.''
\emph{D Magazine}, June.
\url{http://www.dmagazine.com/publications/d-magazine/2014/june/dallas-first-microbreweries.}

Infante, David. 2015. ``The Hipster is Dead, and You Might Not Like Who
Comes Next.'' \emph{Mashable}, June 9.
\url{http://mashable.com/2015/06/09/post-hipster-yuccie/}

Jacobs, Jane. (1961) 1992. \emph{The Death and Life of Great American
Cities}. New York: Vintage Books.

Li, David K. 2014. ``Brooklyn Brewery Plans New Staten Island Plant.''
\emph{New York Post}, July 29.
\url{http://nypost.com/2014/07/29/brooklyn-brewery-plans-new-staten-island-plant/}

Lees, Loretta, Tom Slater, and Elvin Wyly. 2008. \emph{Gentrification}.
New York: Routledge.

Logan, John R., Zengwang Xu, and Brian Stults. 2014. ``Interpolating US
Decennial Census Tract Data from as Early as 1970 to 2010: A
Longitudinal Tract Database.'' \emph{The Professional Geographer} 66
(3): 412--420. doi:10.1080/00330124.2014.905156.

Marroquin, A. 2014. ``Anaheim Streamlines the Permit Process for New
Craft Breweries.'' \emph{The Orange County Register}, March 13.
\url{http://www.ocregister.com/articles/city-605333-county-anaheim.html}

Mathews, Vanessa, and Roger M. Picton. 2014. ``Intoxifying
Gentrification: Brew Pubs and the Geography of Post-Industrial
Heritage.'' \emph{Urban Geography} 35 (3): 337--56.
doi:10.1080/02723638.2014.887298.

May, Lucy, and Dan Monk. 2015. ``Cincinnati Planners Tap Craft Brewers for Help in Making City More Receptive to Booming Industry.'' \emph{WPCO}, May 10. \url{http://www.wcpo.com/entertainment/local-a-e/beer/cincinnati-planners-tap-craft-brewers-for-help-in-making-city-more-receptive-to-booming-industry}

McLaughlin, Ralph B., Neil Reid, and Michael S. Moore. 2014. ``The
Ubiquity of Good Taste: A Spatial Analysis of the Craft Brewing Industry
in the United States.'' In \emph{The Geography of Beer}, edited by Mark
Patterson and Nancy Hoast-Pullen, 131--54. Dodrect: Springer.

Newman, Kathe, and Elvin Wyly. 2006. ``The Right to Stay Put, Revisited:
Gentrification and Resistance to Displacement in New York City.''
\emph{Urban Studies} 43 (1): 23--57. doi:10.1080/00420980500388710.

Ogle, Maureen. 2006. \emph{Ambitious Brew: The Story of American Beer}.
Orlando: Harcourt.

Perritt, Marcia. 2013. ``Breweries and Economic Development: A Case of
Home Brew.'' \emph{Community and Economic Development program at the UNC
School of Government} (blog), April 5.
\url{http://ced.sog.unc.edu/breweries-and-economic-development-a-case-of-home-brew/}

Peters, Corbin, and Zach Szczepaniak. 2013. ``Beer: Is It Zoned Out?''
\emph{Plan Charlotte} (blog), February 6.
\url{http://plancharlotte.org/story/beer-it-zoned-out}

PubQuest. 2015. Accessed April 21. http://www.pubquest.com

Schnell, Steven M., and Joseph F. Reese. 2003. ``Microbreweries as Tools
of Local Identity.'' \emph{Journal of Cultural Geography} 21 (1):
45--69. doi:10.1080/08873630309478266.

---------. 2014. ``Microbreweries, Place, and Identity in the United
States.'' In \emph{The Geography of Beer}, edited by Mark Patterson and
Nancy Hoast-Pullen, 167--87. Dodrect: Springer.

Somerville, H. 2013. ``Oakland: Craft Beer Trend Helps Rebuild
Neighborhoods.'' \emph{San Jose Mercury-News}, July 14.
\url{http://www.mercurynews.com/ci\_23671853/oakland-craft-beer-trend-helps-rebuild-neighborhoods}

Stradling, David, and Richard Stradling. 2008. ``Perceptions of the
Burning River: Deindustrialization and Cleveland's Cuyahoga River.''
\emph{Environmental History} 13 (3): 515--35.

Tremblay, Carol Horton, and Victor J. Tremblay. 2011. ``Recent Economic
Developments in the Import and Craft Segments of the US Brewing
Industry.'' In \emph{The Economics of Beer}, edited by Johan F. M.
Swinnen, 141--60. Oxford; New York: Oxford University Press.

Tremblay, Victor J., and Carol Horton Tremblay. 2005. \emph{The U.S.
Brewing Industry: Data and Economic Analysis}. Cambridge, MA: MIT Press.

United States Census Bureau. 2000. ``Census Regions and Divisions of the
United States.''
\url{http://www2.census.gov/geo/pdfs/maps-data/maps/reference/us\_regdiv.pdf}

Watson, Bart. 2014. ``The Demographics of Craft Beer Lovers.'' Presented
at the Great American Beer Festival, Denver, CO, October 3.
\url{https://www.brewersassociation.org/wp-content/uploads/2014/10/Demographics-of-craft-beer.pdf}

\end{document}